\documentclass[letterpaper,pre,floatfix,superscriptaddress%
,showpacs,twocolumn]{revtex4}
\usepackage{epsfig,graphics,amsmath,amssymb}
\begin{document}
\title{Pattern formation in a metastable, gradient-driven sandpile}
\author{Lucian~Anton}\email{anton@ifin.nipne.ro} 
\affiliation {Institute of Theoretical Physics, 
University of Stellenbosch, Private
Bag X1, 7602 Matieland, South Africa} 
\affiliation{Department of
Physics and Astronomy, University of Manchester, M13 9PL, U.K.}
\affiliation{Institute of Atomic Physics, INFLPR, Lab 22, PO Box MG-36
R76900, Bucharest, Romania}
\author{Hendrik~B.~Geyer}\email{hbg@sun.ac.za} 
\affiliation {Institute of Theoretical Physics, University of
Stellenbosch, Private Bag X1, 7602 Matieland, South Africa} 
\date{} 

\begin{abstract}
With a toppling rule which generates metastable sites, we explore 
the properties of a gradient-driven sandpile that is minimally
perturbed at one boundary. In two dimensions we find that the
transport of grains takes place along deep valleys, generating a
set of patterns as the system approaches the stationary state. We use 
two versions of the toppling rule to analyze the time behavior and the
geometric properties of clusters of valleys, also discussing the
relation between this model and the general properties of models
displaying self-organized criticality.
\end{abstract}


\maketitle

\section{Introduction}

Sandpile models were first introduced by Bak, Tang, and Wiesenfeld as
explicit models of self-organized criticality (SOC)
\cite{Bak87}. Since then a vast literature has analyzed sandpile
properties resulting from various definitions of the toppling rules; 
see Refs.\cite{Mehta96,Jensen98,Dhar99} for recent reviews.

In this paper we present the relaxation and stationary properties of a
gradient-driven sandpile model with a metastable toppling rule.
Interest in this type of work originates from seismology, where
quasiperiodic behavior of seismic activity has been observed for
certain faults and investigated with SOC-related models
\cite{Dahmen98}.  The model we present can describe characteristics of
systems undergoing rheologic flow, e.g., in the mining environment or
tectonic plates.  In such systems stress can accumulate in various
parts of the system for long periods of time, in contrast to normal
fluid flow in which the local relaxation time is much smaller than the
hydrodynamic time scale.

Furthermore, we believe that the properties we describe are of wider
interest to other fields of nonequilibrium statistical mechanics
where, e.g., stripe patterns similar to those that we observe appear
in a variety of extended systems, including sand and biological
systems\cite{Ball99,Shinbrot01}. Our approach is quite general since
it utilizes only a consistent local ordering of topplings, following
the instantaneous maximum gradient, and the notion of a metastable
site. Furthermore, the model we propose shows a quasiperiodic time
behavior, a feature already mentioned for SOC-related models in
\cite{Socolar93} and explored recently in a similar context by
Chapman\cite{Chapman00}.

The additional feature of the proposed model is the emergence of a
spatial structure for the metastable configuration driven by the
transport of grains through the system. A study of this spatial
configuration in and close to the stationary state is the main
objective of the following sections.

The paper is organized as follows. In Sec. II we describe the toppling
rule and its connection with related models. In Sec. III we present
the 1D variant of the model and in Sec. IV the results for the 2D
version, followed by conclusions in Sec. V.

\section{The toppling rule}

We consider a gradient-driven sandpile with a toppling rule which
takes into account not only whether a local threshold gradient is
exceeded, but also whether this situation is the result of the {\it
addition} of a grain to the site under evaluation. We introduce this
rule as a simple description of the dynamical weakening introduced in
Ref.\cite{Dahmen98}.

The general description of the dynamics is as follows. Grains are
dropped on the sites in a designated region of the lattice called the
source zone; following a relaxation rule, to be described in detail
subsequently, grains can change their position on the lattice until
they reach an open boundary and leave the system (lattice).  In short,
the model describes the transport of grains from the source zone to
the open boundary.

Let us now analyze in detail the toppling rule in 2D on a square
lattice.  At a given moment of time each site of the lattice has an
associated height $h$ of the grain column, and we also associate with
it the set of gradients $G=\{g_l,g_r,g_u,g_d\}$, where
$g_\alpha=h-h_\alpha$ and $h_\alpha\in \{h_l,h_r,h_u,h_d\}$ is the the
height of the sand column at the four nearest neighbor sites left,
right, up, and down.

We use two thresholds in our algorithm. (i)$g_{max}$ is the stability
threshold for the maximum of $G$.  If at least one of the gradients of
$G$ is larger than $g_{max}$ the site is called metastable.  It
topples only if this state is the result of the receipt of a grain,
either from a neighboring site which had toppled, or when a grain was
dropped on the site at the start of an updating run. Metastable states
do {\it not} topple when a gradient larger than $g_{max}$ develops as
a result of the loss of grains on neighbouring sites. (ii) $g_{min}$
is the minimum positive gradient, which fixes the condition to stop
the toppling; we call it the activity threshold.  Once a site starts
toppling, it sends grains, one at a time, along the instantaneous
maximum gradient of $G$ to its nearest neighbors, provided that
max$\{G\}>g_{min}$.  If there is more than one instantaneous maximum
gradient a random choice is made among them. If one site topples it
will send grains to some of its nearest neighbors. We refer to the
process of relaxation of one site as a ''toppling'' and to the sites
that have received sand grains during toppling as ''updated sites''.
Their coordinates are kept in a list for the next step of the
dynamics.

In one time step all the updated sites produced at the previous time
step are checked for stability and, if unstable, they relax according
to the above algorithm. We specify that once a site is unstable and
chosen to topple the algorithm will finish the toppling sequence at
the site and then will move to another site. Physically, this is
equivalent with neglecting the toppling time.

Because we always topple along the instantaneous maximum gradient we
can introduce a time order of the topplings.  On physical grounds it
seems reasonable to assume that the site that first receives a
grain will topple first at the next time step, if it is unstable.  We
model this using a first in, first out list: the coordinates of the
updated sites are stored sequentially in a list. In one time step the
algorithm reads the list sequentially from the first entry: if the
current site is metastable, it is toppled and the sites that are
updated in this process are stored in the list for the next time step;
if the current site is stable, it is simply discarded from the list.  We
refer to this time ordered toppling rule by the acronym TOTR.  (Recall
that a gradient toppling rule implies a non-Abelian sandpile where the
order of toppling has definite consequences; see, e.g.,\
Refs.\cite{Jensen98,Dhar99}.)

Alternatively, we can disregard the time ordering and select at random
a site from the list of updated sites for inspection of its stability.
In Sec. III we make a comparative study of these two variants of
the toppling rules in the 2D case.  We refer to this randomly ordered
toppling rule by the acronym ROTR.

In other words, we can consider one time step of the dynamics, $\Delta
t$, divided into $N$ bins ($N$ being the number of sites of the
lattice), each bin containing at most the coordinates of one updated
site since in a time interval $\Delta t/N$ we can assume that
typically at most one site is active.  An active site generates a set
of updated sites when topples. In this description the TOTR fills the
bins of the next time step in the order in which the updated sites are
produced at the current time step; meanwhile the ROTR fills the bins
of the next time step randomly. Evidently, most of the bins are empty
as the number of updated sites is much smaller than the system size
and the algorithm discards the empty bins using the list of updated
sites.

Thus, to resume, at a given instant of time we have three kinds of
sites in our system: (i) inactive sites with all associated gradients
less than $g_{max}$; (ii) metastable sites, which have at least one
associated gradient larger than $g_{max}$; and (iii) active or
unstable sites, which are the toppling sites at the given instant
(they are perturbed metastable sites).

One observation about the times scales in the model is now in order.
We can think in terms of the existence of three time scales. The
smallest one is the time in which a site has toppled, $\tau_t$. The
second time scale is the surviving time of an updated unstable site,
$\tau_s$. The dynamics of our model is such that
$\tau_s\gg\tau_t$. The third time scale is the time between two
droppings of grains on the lattice $\tau_a$. After one grain is
dropped the system relaxes globally through an avalanche of
topplings. Here we are ensured that $\tau_a\gg N_{L}\tau_s$, where
$N_{L}$ is the average number of time steps for the system to relax
for a given size $L$.

We define the size of an avalanche as the total number of
topplings generated in a given relaxation process after one dropping.

We stress that our toppling rule allows and introduces sites with
unstable gradients after one avalanche has taken place.  They emerge
as the neighbors of the toppling (active) sites toward which the
gradient is negative.  As an unstable site topples, the negative
gradient increases in absolute value, but the sites along such a
direction {\it do not receive any grains} and hence are not toppled,
according to our rule.  When topplings in such an avalanche stop, such
a site can accordingly have a maximum gradient that is larger than
the threshold value $g_{max}$.  We call such a site metastable --- it
can sustain gradients larger than the threshold as long as it remains
unperturbed.  Physically, we associate this rule with a certain local
metastability of the medium through which transport takes place.

We make the observation that the metastable sites may be important in
the regions where there is no dropping of grains, since they can live
much longer than the average time between two droppings on the same site.

As elaborated in the next two sections, this model has properties
similar to the extended version of the forest fire model described in
Ref.\cite{Socolar93}, that is, quasiperiodic behavior in time and a
spiked avalanche distribution.

To link further with previous studies, it is also of interest to
understand the relation between our proposed model and the condition
for SOC in sandpile models.  References \cite{Socolar93,Grinstein90}
present an analysis to establish the generic conditions a model has to
satisfy in order to present SOC. The authors show that in two
dimensions, or larger, and for conservative dynamics, an intrinsic
spatial anisotropy is required to produce SOC for models which can be
treated perturbatively. The model we propose has a conservative
toppling rule (the noise is also conservative in the region where no
dropping takes place) and evolves in two spatial dimensions under
conditions of anisotropy, but it does not present the features of SOC for
the avalanche distribution.

We think that the explanation of this anomaly resides in the fact that
the stationary state of this model cannot be characterised as a
perturbation of an interaction-free model. As we present in full detail
in Sec. \ref{2d}, the stationary state of this model is
characterized by the appearance of deep and narrow valleys along the
direction of grain transport. These features cannot be obtained as a
perturbation of a diffusionlike relaxation.

We also make the observation that in the proposed toppling rules there
is no external noise term, e.g.,\ similar to the noise term in the
Langevin equation. The randomness in our model comes from the
selection of the toppling order in the case of the ROTR, and from the
random choice of the toppling direction in the case of a degenerate
maximum gradient. Another source of randomness in these models is the
`internal noise', in the sense discussed in Ref. \cite{vanKampen92},
coming from the fluctuations of the many particle dynamics which may
be present even in the case of pure deterministic microscopic
equations. There is no straightforward connexion between the `internal
noise' and the `external noise' term of a Langevin equation.

In connection with this theoretical aspect we mention that recently a
model with metastable states (''sticky grains'' is the term used by
the authors) but with a height-driven toppling rule and dissipative
dynamics was studied in Ref. \cite{Mohanty02} and shown to be in the
directed percolation universality class.

In the following sections we present a detailed discussion of the 1D
case and a statistical analysis of the mentioned patterns of
metastable states for the 2D case in the stationary state and close to
the stationary state.

\section{The 1D case}

We start our study with a one-dimensional sandpile.  We choose a
lattice of dimension $L$ with an open boundary at $x=L$ and with a
wall at $x=0$.  The grains are injected randomly in the region $x \in
[1,\; w]$, with $1\le w \le L $, called the source zone. We choose the
stability threshold $g_{max}=2$ and the activity threshold
$g_{min}=1$.  Let us analyze in detail the appearance of a metastable
site in 1D lattice.  For simplicity we start with an initial stair
configuration which has a monotonic step of max$\{G\}=2$ descending
from $x=1$ to $x=L$.  This configuration is marginally stable; if we
drop a grain anywhere on the lattice an avalanche occurs.  First we
consider a restricted source zone where grains are only dropped at
$x=1$.  A grain dropped at $x=1$ will then topple until $x=L$ for the
chosen configuration.  At $x=L$ an extra toppling takes place since
$g_{min}=1$, and the toppling of the disturbed state continues as long
as the maximum gradient is larger than $g_{min}$.  [Formally, the
boundary condition is that $h(L+1)=0$, $h(0)=\infty$.]  When the
avalanche stops, we therefore see that the gradient between the sites
$L-1$ and $L$ is $3$, hence we have a metastable site at $L-1$.  Now
if we drop a new grain at $x=1$, the same consideration shows that the
metastable configuration moves to $L-2$.  The process continues until
the metastable configuration reaches the site $x=1$.  After that a
series of avalanches reconstructs the initial configuration, adding a
grain at site $L$, then at $L-1$, and so on.

\begin{figure}[t]
\begin{minipage}[b]{0.95\linewidth}
\epsfig{file=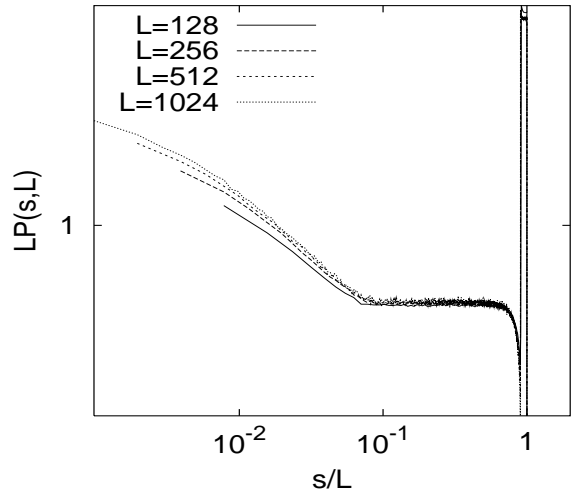,height=\linewidth,width=%
\linewidth,clip=,bbllx=50pt,bblly=120pt,bburx=550pt,bbury=700pt}
\caption{\label{1D} The avalanche size distribution in $1$D at
  $L=128$, $256$, $512$, and $1024$ with the fixed ratio $w/L=0.1$. We
  notice the peak in the top right corner of the plot.}
\end{minipage}
\end{figure}

The characteristic time period of the system is controlled by the time
in which the metastable site travels from the site $L$ to the site
$1$. This kind of behavior is preserved if we use a source zone with
$w>1$ but with the ratio $w/L$ small. The avalanche size distribution
is characterized by a peak at the end of the distribution support (see
Fig. \ref{1D}), which results from the avalanches produced while the
metastable site is present in the transport zone between $w+1$ and
$L$. The smaller size avalanches are produced after the metastable
site has reached the source zone.  Figure \ref{1D} shows  that for
fixed ration $w/L$ the avalanche distribution scales with $L$.

We close this section with the observation that the distinction
between the TOTR and ROTR is irrelevant in $1$D since typically
there is only one updated site in the algorithm.

\section{The 2D case}\label{2d}

\subsection{General features}

In this section we compare the behavior of the system in two
dimensions with the simple and well understood behavior in 1D.  We
choose for study a rectangular lattice of size $L$ with an open
boundary condition at $x=L$ and a wall at $x=1$, with a periodic
boundary condition along the $y$ axis.  A simulation starts with the
initial condition specified by a uniform slope of size $1$ per lattice
step along the $x$ axis, the height at $x=L$ being $0$.  The initial
slope along the $y$ axis is set to $0$.  We choose the stability
threshold $g_{max}=4$ and the activity threshold $g_{min}=1$.  (We
have also performed simulations starting with an empty lattice; the
stationary state is the same and only the transient regime is longer.)
We use both toppling rules: (i) with the time ordered toppling rule
(TOTR) and (ii) with a random ordered toppling rule (ROTR).

\begin{figure}[t]
\begin{minipage}[b]{0.95\linewidth}
\centering \epsfig{file=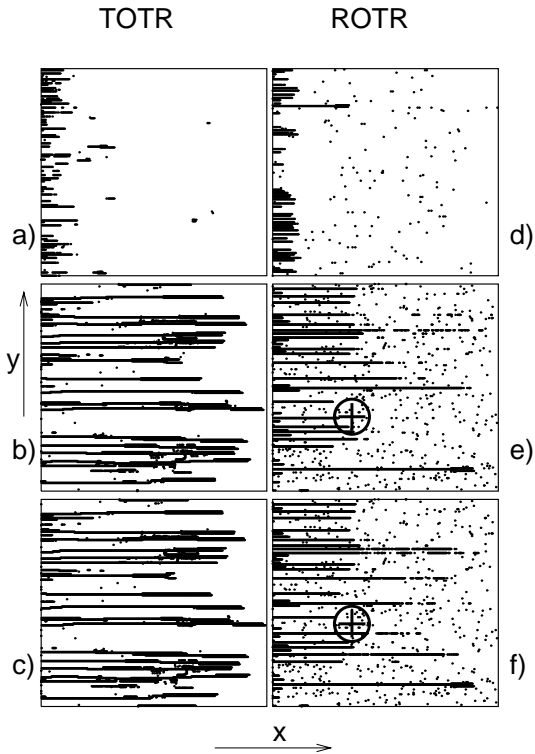, height=1.3\linewidth,
width=\linewidth,clip=,bbllx=60pt,bblly=50pt,bburx=600pt,bbury=750pt}
\caption{\label{ripple} The patterns created by sites at the bottom of
valleys for a $512\times 512$ lattice during the transient regime
(a),(d) and in the stationary regime (b), (c), (e), (f). On the left
side the dynamics is TOTR and on the right side the dynamics is ROTR.
The dots, often merged to lines, mark the sites which have at least
two negative gradients larger in absolute value than $g_{\rm max}$,
i.e., at least two neighbors are metastable. (See the text for
discussion of the symbols $\oplus$.)}
\end{minipage}
\end{figure}

The grains are dropped randomly at $x=1$, $y\in [1, L]$. This source
configuration allows us to explore the behavior of the transport
region of the sandpile under minimal perturbation, since only one
grain topples at the boundary of the transport region and the
metastable sites do not have their average lifetime constrained by the
average time between two droppings on the same site; hence the
structure of the transport zone is influenced only by a minimal flow
of grains.

The initial condition we start with can be viewed as a collection of
interacting 1D sandpiles oriented along the $x$ axis, and consequently
one might expect to see two-dimensional avalanches created by the
interaction of different metastable sites.

\begin{figure}
\begin{minipage}[b]{0.95\linewidth}
\epsfig{file=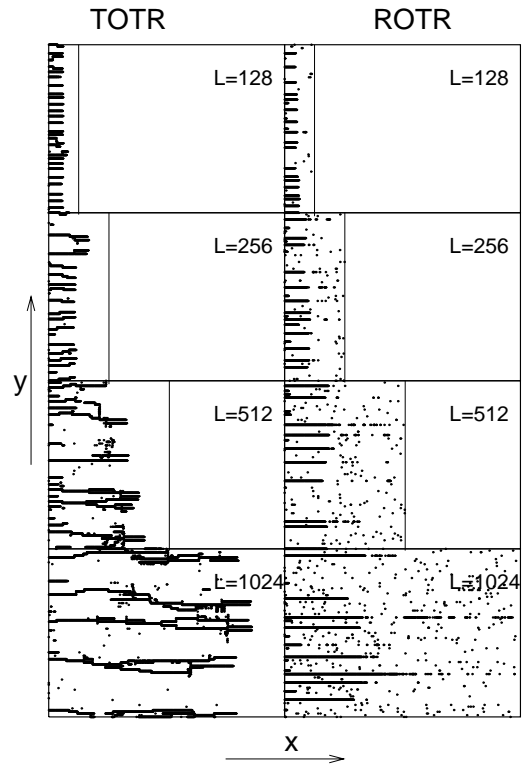,
height=1.3\linewidth,
width=\linewidth,clip=,bbllx=50pt,bblly=60pt,bburx=580pt,bbury=750pt}
\caption{\label{crossover}The structure of the bottom of the valleys
as a function of the lattice size $L$.  From top to bottom $L=128,
256, 512, 1024$; TOTR on the left side and ROTR on the right side. The
source is on the left side and the open boundary on the right side of
each panel. The vertical lines mark the open boundary.}
\end{minipage}
\end{figure}

Surprisingly we found that in the stationary state the sandpile
develops a structure of valleys along the $x$ direction separated by
terraces of sites in the metastable state. As one can see from
Fig. \ref{ripple} the valleys are not purely $1$D; instead they
fluctuate slightly along the transverse direction and also show
branched structures. Visual inspection of Fig. \ref{ripple} shows that
the transverse fluctuations and branching are more pronounced in the
case of the TOTR algorithm.

After the system reaches the stationary state we notice that in the
case of the TOTR the valleys do not change in time except for small
fluctuations which appear toward the open boundary. In the case of the
ROTR the valleys do change in time even in the stationary state. We
illustrate this in Fig. \ref{ripple} where the snapshots (c) and (f)
follow after $10^6$ droppings on the snapshots (b) and (e) in the
stationary regime. We observe that in the case of the TOTR the two
configurations are almost identical, while for the ROTR there is a
clear difference between the two configurations around the points
denoted with the symbol $\oplus$. We return to this point when we
analyze the cluster size distribution and the correlation length
behavior.

\begin{figure}
\begin{minipage}[b]{0.95\linewidth}
\epsfig{file=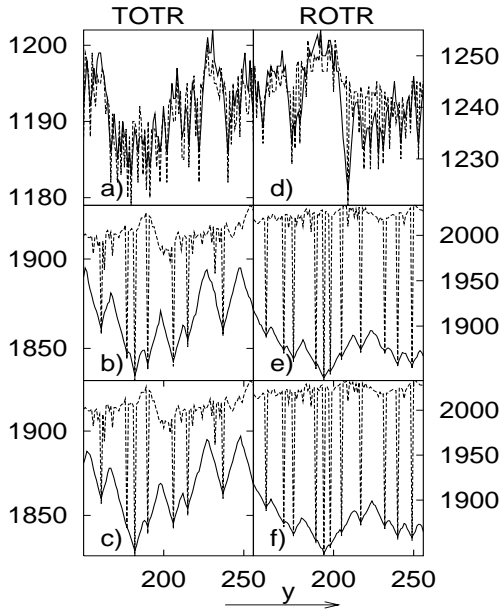, height=1.1\linewidth,
width=\linewidth,clip=,bbllx=0pt,bblly=50pt,bburx=552pt,bbury=750pt}
\caption{\label{profile} The profile of the sandpile along the source
line $x=1$ (continuous line) and along the line next to it, $x=2$
(dashed line), for TOTR (left) and ROTR (right). We show one moment in
time of the transient regime. (a) and (d) , while (b), (c) and (e),
(f) are in the stationary regime. The coordinate is the height in
number of grains. Note the change of the profile height with time, and
the fact that the profile is unchanged in the stationary regime.}
\end{minipage}
\end{figure}

Another observation which we can make from visual inspection of the
stationary patterns, is that the structure of the valleys changes with
system size.  From Fig. \ref{crossover} we see that for $L=128$ the
valleys are predominantly onedimensional.  When the lattice size
increases, transverse fluctuation and branches appear.

\begin{figure}
\begin{minipage}[b]{0.95\linewidth}
\epsfig{file=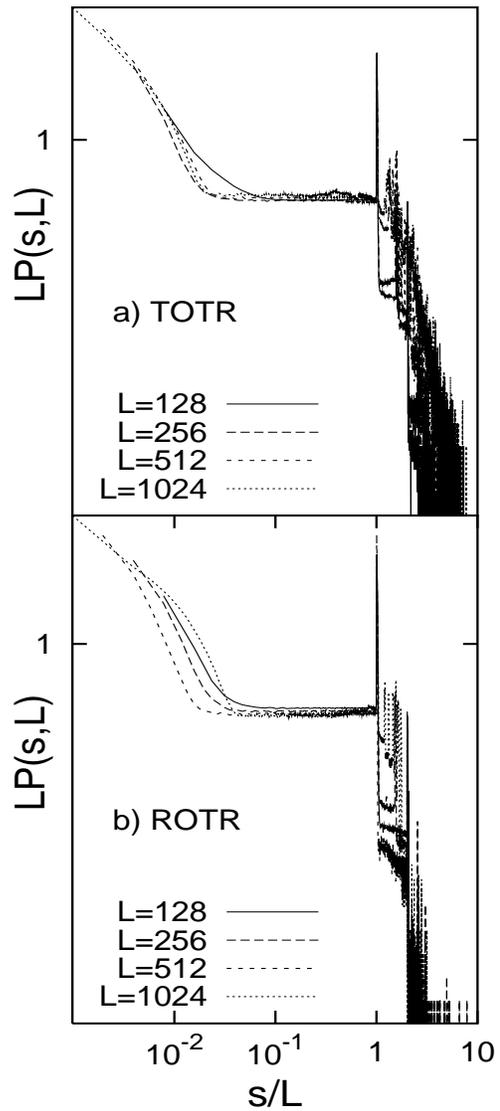, height=2.05\linewidth,
width=\linewidth,clip=,bbllx=150pt,bblly=70pt,bburx=420pt,bbury=700pt}
\caption{\label{av-dist} The avalanche size distribution: (a) TOTR, (b)
ROTR.  The system size takes the values $L=128, 256, 512, 1024$. The
plateau region and the first peak scale approximately with system
size $L$.}
\end{minipage}
\end{figure}

We obtain further insight about the nature of the stationary state if
we analyze the time evolution of the height profile along $z$
(perpendicular to the $x-y$ plane) at the boundary of the source zone.
In Fig. \ref{profile} we show the time evolution of the height profile
along $y$ for $x=1$ and $x=2$.  We see that in the transient state,
for both toppling rules, the source zone ($x=1$) has heights close to
the heights of the $x=2$ profile.  In this configuration a grain dropped
in the source zone will move more probably along the $x$ axis, leaving
the source zone in one or a few steps.  In contrast, in the
stationary regime we see that a completely different configuration
arises.  The average height of the source zone is significantly
smaller than the average height of the $x=2$ profile.  A grain dropped
in the source zone will therefore first travel along the $y$
direction, until it meets a minimum which is connected with a valley,
see Figs. \ref{profile} (b), and (e). In the stationary state the profile at
the source remain unchanged for both the TOTR and ROTR as
Figs. \ref{profile} (b), (c), (e), and (f) show. The data are taken from the
same configuration presented in Fig. \ref{ripple}.

We close the presentation of general features of the model with an
analysis of the avalanche size distribution in the stationary state.
Figure \ref{av-dist} shows for both dynamics that the size
distribution is not power-law-like, but rather similar to the $1$D
case (compare Fig. \ref{1D}), although with a richer structure of
peaks. We remark that a scaling proportional with lattice size $L$
holds approximately for the plateau region and the first peak in the
distribution, as in the $1$D case. Since the valley length scales with
$L$, we think that this feature of the avalanche size distribution is
determined by the propagation of the avalanches along the valleys.

\subsection{Cluster characterization}

Having presented a visual description of the valleys in the previous
subsection, we now consider a quantitative approach.  For a numerical
description of the patterns shown in Figs.  \ref{ripple} and
\ref{crossover} we concentrate on the clusters formed by the bottom of
the valleys.  We define a site to be at the bottom of a valley if it
has associated with it two or more negative gradients larger in
absolute value than the threshold, i.e., at least two neighbors are
metastable.

A set of bottom of the valley sites is said to form a cluster if they
can be spanned by a path stepping only to nearest neighbors along the
$x$ or $y$ directions.

We analyze the following quantities: the total number of clusters
$N_c(t)$ as a function of time, the total cluster mass $M_c(t)$ as a
function of time, the longitudinal and transverse correlation
lengths $\xi_{\parallel}$ and $\xi_{\perp}$, and, finally,  the cluster 
size distribution  at a given time, $N(s,t)$.

We start with the time evolution of the total number of clusters
$N_c(t)$ and the total mass $M_c(t)=\sum_s sN(s,t)$.  We remind the
reader that in this section a time step corresponds to a grain drop
and the subsequent relaxation, and therefore we neglect the relaxation
time per avalanche.

\begin{figure}
\begin{minipage}[b]{0.95\linewidth}
\epsfig{file=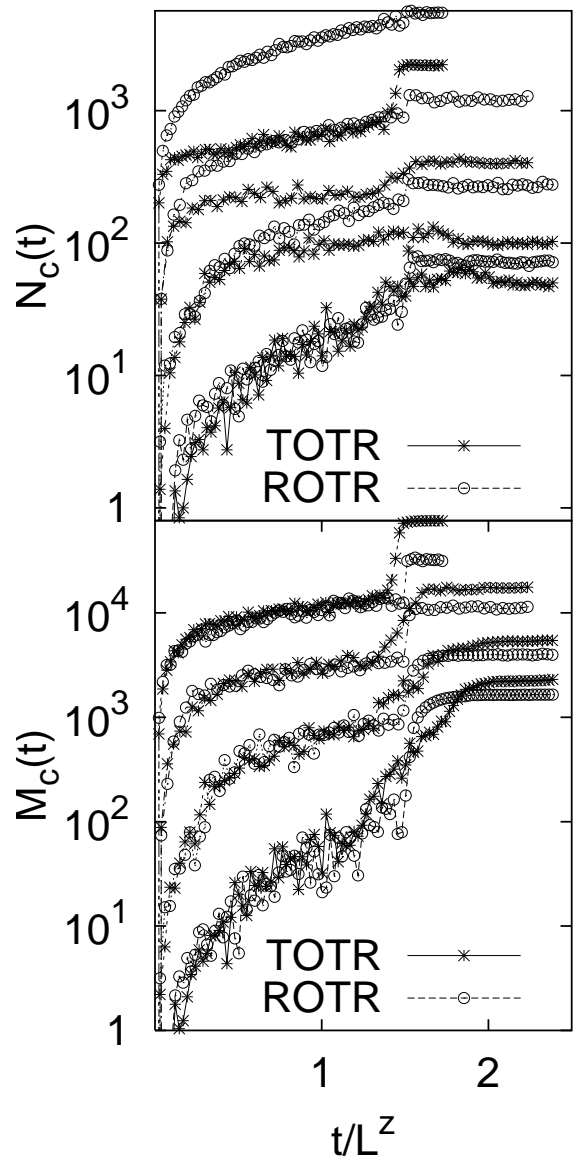,height=1.9\linewidth,
width=\linewidth,clip=,bbllx=150pt,bblly=85pt,bburx=500pt,bbury=750pt}
\caption{\label{ncoft} Number of clusters (top) and their total
  mass (bottom) as function of time for the TOTR and ROTR. In
  ascending order the plots correspond to the lattice sizes
  $L=128$, $256$, $512$, $1024$.}
\end{minipage}
\end{figure}

The data were collected over a single run starting with the initial
condition specified at the beginning of this section. We have collected
the data for one point at time intervals $5\times 10^4$, $10^5$, $10^6$,
and $5\times 10^6$ for the system sizes $L=128$, $256$, $512$, $1024$,
respectively. To eliminate the short time fluctuation we averaged
at each point over a window of size $10^4$ steps from which we
selected $100$ moments equally spaced. We mention here that we are
constrained to use single runs since the characteristic time in which the
system reaches the stationary state scales approximately with the
third power of the lattice linear size.

Figure \ref{ncoft} shows that the total mass of the clusters and the
number of clusters have a jump before the stationary state is
reached. The jump is more pronounced in the case of the TOTR and is
clearer for large system size. We observe that the time to reach the
stationary state scales with the third power of system size. This fact
can be explained if we assume that the average slope of the system
converges to a nonzero value as the system size is increased; then the
volume of the accumulated grains scales like $L^3$, which determines
the minimal number of time steps necessary to drive the system to the
stationary state.

In studies of this kind of model it is customary to test the data for
scaling behaviour $ f(t,L)=L^{\alpha}\tilde f(t/L^{z})$. Our plot
shows that one can define a dynamical exponent $z=3$ for the
relaxation time, but a simple proportionality with the system size at
a given power is not sustainable, although for $L=512, 1024$ the
assumption holds acceptably (the corresponding lines are parallel
within a good approximation). However, the scaling hypothesis does not
hold for the system size $L=128$, suggesting that boundary effects
have a characteristic length of this order of magnitude. From
Fig. \ref{crossover} we see that close to the the source zone the
valleys do not present branching. We note also that close to the open
boundary valleys do not form, which makes it plausible that the
scaling is strongly affected by finite size effects.

\begin{figure}
\begin{minipage}[b]{0.95\linewidth}
\epsfig{file=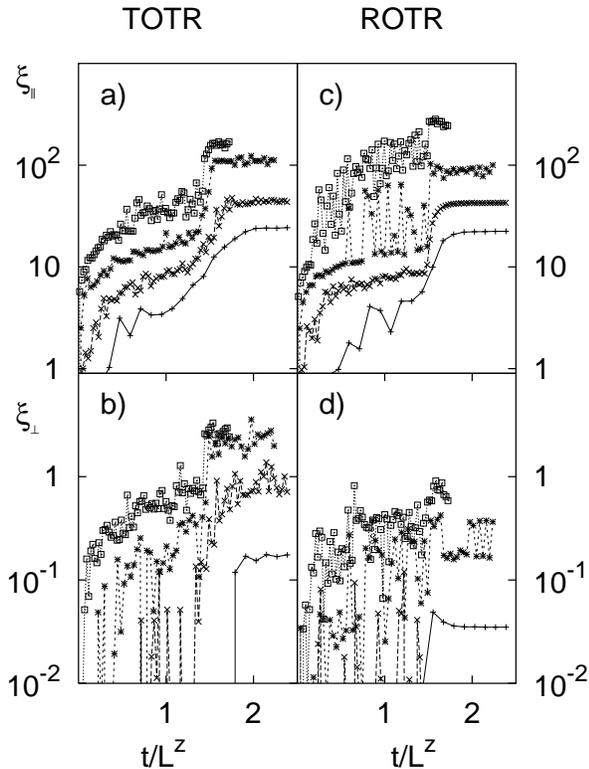,
height=1.35\linewidth,
width=\linewidth,clip=,
bbllx=50pt,bblly=50pt,bburx=550pt,bbury=730pt}
\caption{\label{corrlenT} Correlation lengths $\xi_\parallel$ (top)
  and $\xi_\perp$ (bottom) as function of time [ L=$128$ ($+$),
  $256$($\times$), $512(*)$, $1024(\Box)$]. Left panels are for TOTR
  right panels for ROTR.}
\end{minipage}
\end{figure}

For a geometric charaterization of the clusters we define the
following correlation lengths, following Ref.  \cite{Stauffer92}:
\begin{equation}\label{eq:xi}
\xi^{2}_{i}=\frac{2\sum_s R^{2}_{i,s}s^2n_s}{\sum_s s^2n_s},
\end{equation}
where $R^{2}_{i,s}$ is the mean square distance along the
longitudinal, $x$, or transverse, $y$, direction for the sites
belonging to one cluster averaged over all clusters of size $s$, which
can be expressed in the formula
\begin{equation}\label{eq:sqdisp}
2R^{2}_{i,s}=\frac{1}{n_s}\sum_{\alpha=1}^{n_s} \frac{1}{s^2}
\sum_{m_\alpha,n_\alpha}|x^{(i)}_{m_\alpha}-x^{(i)}_{n_\alpha}|^2 \; .
\end{equation}
Here the index $i=\parallel,\perp$ indicates the longitudinal and
transverse directions, $x^{(i)}_{n_\alpha}$ denotes the coordinate of
the sites belonging to the $\alpha$th cluster of size $s$, and $n_s$
is the number of clusters with size $s$. In other words, for each
cluster from a given configuration of the lattice we compute its
average square displacement from the center of mass in transverse and
longitudinal directions with Eq. (\ref{eq:sqdisp}). Next we average
over all the clusters from a configuration with weights specified by
Eq. (\ref{eq:xi}).

Figure \ref{corrlenT} shows that the longitudinal correlation lengths
jump before the stationary regime for system sizes $L=512$ and $1024$, 
similar to the behavior of the cluster number and the total mass( see
Fig. \ref{ncoft}). 

It is intriguing that the longitudinal correlation length shows
chaotic behavior in the transient state as the system size increases,
especially in the case of the ROTR, even though we have averaged out
the short time fluctuations.  On the other hand, the time evolution of
the total number of clusters does not present fluctuations of the same
relative magnitude (see Fig. \ref{ncoft}). We conjecture that the
large fluctuation of the correlation lengths in the case of the ROTR is
related to a high frequency of coalescence and breaking up of
clusters before the stationary state is reached.

\begin{figure}
\begin{minipage}[b]{0.95\linewidth}
\epsfig{file=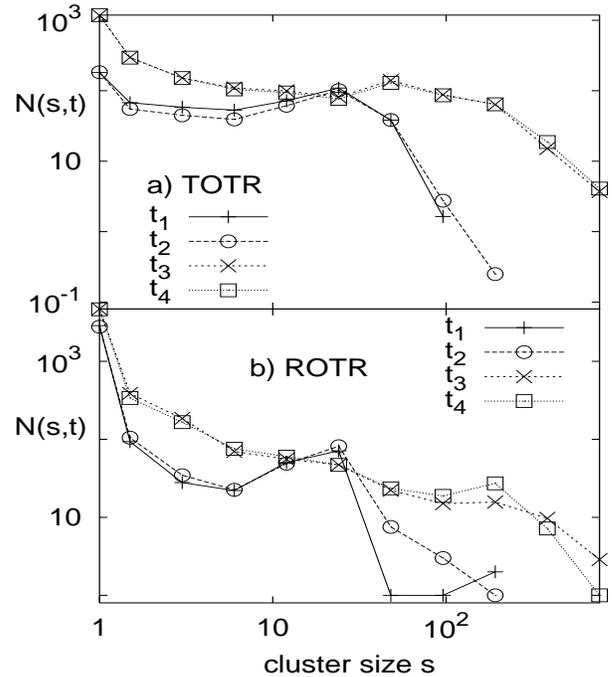,
height=1.4\linewidth, width=\linewidth,clip=,
bbllx=50pt,bblly=50pt,bburx=550pt,bbury=900pt}
\caption{\label{histogram}Histogram of cluster sizes at four moments
of time. We take $t_1$ and $t_2$ in the transient regime with
$\xi_\parallel(t_1)<\xi_\parallel(t_2)$ and $t_3$, $t_4$ in the
stationary regime with $\xi_\parallel(t_3)<\xi_\parallel(t_4)$. We
mention that data are averaged around each time moment $t_i$,
$i\in[1,4]$, as described at the beginning of Sec. IV B.}
\end{minipage}
\end{figure}

The transverse correlation presents the same kind of chaotic
behavior. In any event, we see that $\xi_\perp \ge 1$ (which is
physically significant) only for the TOTR with $L\ge 512$. Visual
inspection of Figs. \ref{ripple}, and \ref{crossover} shows that it is
this situation in which a significant number of branchings are
observable.

To elucidate this point further, we study, for both types of dynamics
TOTR and ROTR, the cluster size distribution at various moments of
time for the lattice size $L=1024$. In Fig. \ref{histogram} we plot
the cluster distribution at two consecutive moments in the transient
regime ($t_1=8.7\times 10^8$, $t_2=8.75\times 10^8$) corresponding to
a minimum and a maximum of $\xi_\parallel$ observed in
Fig. \ref{corrlenT}, that is,
$\xi_{\parallel}(t_1)<\xi_{\parallel}(t_2)$, and in the stationary
state ($t_3=1.84\times 10^9$, $t_4=1.845\times 10^9$) with the same
condition $\xi_{\parallel}(t_3)<\xi_{\parallel}(t_4)$ .

The plots show that in the transient regime the states with larger
$\xi_{\parallel}$ have a significant population of large clusters for
both dynamics. In the stationary state in the case of the TOTR the
cluster distribution changes very little, while for the ROTR the the
large size tail presents observable fluctuations. This observation
seems to point to a signature for a stationary state of the ROTR in
which large scale fluctuations appear a the cluster of valleys, as
shown in Fig. \ref{ripple}. (We have also checked this fact for other
configurations.)

To summarize the two-dimensional behavior, we have studied the
geometric and relaxation properties of the transport region of the
sandpile. We found that as the system approaches the stationary state,
clusters of deep valleys appear. Specifically, we explored the time
evolution for the total number of clusters, the total mass of the
clusters, the correlation lengths (longitudinal and transverse) of the
clusters, and the cluster size distribution at various moments of
time. The main distinction between the TOTR and ROTR is the fact that
the configurations obtained from the ROTR are strongly affected by
fluctuations of the longitudinal correlation length, which we have
shown to be associated with fluctuation in the tail of the cluster
size distribution.

The characteristic time to reach the stationary state scales with the
system size as $L^{z}$, with $z\approx 3$, but the magnitude of the
observed quantities does not follow a clear power law type of
scaling. The range of data we have does not allow us to decide if this
is due to strong correction from boundary effects, or if scaling is
genuinely broken. The setting in of a chaotic regime at large $L$
complicates the analysis even further.

\section{Conclusions}

We have analyzed a gradient-driven sandpile model with local
metastable states using two variants of the toppling rule: one which
keeps the time order of the list of updated sites and one which
selects a site from the list at random. We found that in 2D metastable
sites generate a rough landscape with deep valleys along which grains
are transported.

The valleys are organized in clusters with a pronounced development
along the direction of grain flow ($\xi_{\parallel}\gg\xi_{\perp}$).
For the time ordered toppling rule, we found that transverse
correlations develop as the system size increases. Also, for large
system size we observed chaotic behavior for the correlation lengths,
associated with fluctuations in the size distribution of the clusters.

The avalanches produced in the resulting valleys have one-dimensional
properties and show non-SOC behavior.  We believe that this is an
example of a model in the strong coupling regime which escapes the
classification made in Ref. \cite{Grinstein90} for models that can be
treated perturbatively. Yet we do not have an exact mapping between
our discrete model and the continuum Langevin equations used in
Ref. \cite{Grinstein90}. This may be another source of the
discrepancy. At the moment this is a purely numerical study of a
descriptive nature. To clarify the previous uncertainties further
analytical insight seems to be required. It would, e.g.,\ be interesting
to deduce a simple mean field kind of equation capable of explaining the
irregular time behavior at large $L$ for the quantities described in
the paper.

\section*{Acknowledgments}

The authors thank an anonymous referee for detailed comments and
useful suggestions.  One of the authors, L~A, thankfully acknowledges
the hospitality of the Department of Physics, Inha University, South
Korea, where part of the work was done; L. A. also acknowledges
partial support from the Marie Curie Fellowship of the European
Community programme under Contract No. HPMF-CT-2002-01910 . We
acknowledge early discussions with Morne Pistorius and preliminary
calculations by him which stimulated the present investigation.

\bibliographystyle{apsrev}
\bibliography{/thuser/lucian/papers/biblio}
\end{document}